\documentclass[preprint,showpacs,preprintnumbers,amsmath,amssymb]{revtex4} 

\usepackage{graphicx,bm}

\newcommand{\bbeta}{\bm\beta}
\newcommand{\bnabla}{\bm\nabla}
\newcommand{\vvec}{\bm {\mathit v}}

\begin{document}
\title{Solution of the Multicomponent Boltzmann Equation Based on an Extended Set of Observables}
\author{S.V. Savenko}
\author{E.A.J.F. Peters}
\author{P.J.A.M. Kerkhof}
\affiliation{Transport Phenomena Group, Department of Chemical Engineering \& Chemistry,
						 Eindhoven University of Technology, the Netherlands}
\date{\today}

\begin{abstract}
We present the perturbative  solution  of  the  multicomponent  Boltzmann
kinetic  equation  based  on  the  set  of  observables   including   the
hydrodynamic  velocity  and temperature for each component.
The solution is obtained by modifying the formal density scaling
scheme by Enskog, such that the  density  of  each  component  is  scaled
independently.
As a result we obtain the species momentum and energy  balance  equations
with the source terms describing the transfer of corresponding quantities
between different components.
In the zero order approximation those are the Euler  equations  with  the
momentum and heat  diffusion  included  in  the  form  of  the  classical
Maxwell-Stefan diffusion terms.
The first order approximation results in  equations  of  a  Navier-Stokes
type with the partial viscosity and heat conductivity including only  the
correlations of the particles of the same component.
The first order corrections to the Maxwell-Stefan terms as  well  as  the
contributions bilinear in gradients and differences  of  observables  are
calculated.
The first  order  momentum  source  term  is  shown  to  include  thermal
diffusion.
The nondiagonal (in component indexes) components of viscosity and heat 
conductivity appear as second order contributions.
\end{abstract}
\maketitle

\section{Introduction}
The theory of transport phenomena in gases  based  on  a  kinetic  theory
dates back to the works of Chapman and Enskog
\cite{chapman_1917,enskog_1917,ferziger_kaper_1972}  and   is   generally
accepted as mature.
It provides the perturbative solution of Boltzmann  kinetic  equation  for
the single- as well as multicomponent gases, resulting  apart  from  other
things in the microscopic basis for the Navier-Stokes equations.
In spite of that the applicability of  Chapman-Enskog  solution  to  some
particular problems is limited and a number of works has  been  published
on modified and alternative derivation of transport  equations  from  the
microscopic kinetics
\cite{grad_1963,alavi_snider_1998,chen_spiegel_2001,kerkhof_geboers_2005}.
Dynamics  of  multicomponent  flows  has  also   attracted   considerable
interest  of the statistical mechanical community in the past.
The works on the subject include the derivations of species momentum  and
heat balance equations \cite{bearman_kirkwood_1958,snell_aranow_1967}, as
well as the works devoted to discussion  and  modification  of  diffusion
equations
\cite{krishna_wesselingh_1997,felderhof_2003,runstedler_2006,schimpf_semenov_2004}.
A recent overview of multicomponent transport theories can  be  found  in
\cite{kerkhof_geboers_2005(1),young_todd_2005}.

The common recipe in modeling  $n$-component  transport  problems  is  to
solve the Navier-Stokes equation for the mixture, combined  with  $(n-1)$
Maxwell-Stefan  equations,  all  with  appropriate  boundary  conditions.
Application of this principle to binary counterdiffusion  in  capillaries
\cite{remick_geankoplis_1974} leads however to large  discrepancies  with
experimental  results (see for example \cite{kerkhof_geboers_2005}).
On the other hand, instead of the set of observables including the number
density of each component ($n_i$), the hydrodynamic velocity  ($\vvec_0$)
and the temperature ($T$) of the mixture one may  wish  to  use  the  set
including the velocities and temperatures for every component 
$(\vvec_{i0}, T_i)$. 
This choice of observables was made already in the works of  Maxwell  and
Stefan \cite{stefan_1871,maxwell_1860}, but this has become  obscured  by
the success of the Chapman-Enskog theory.
The species momentum and heat balance equations resulting  from  such  an
approach can be also considered as a basis for multifluid models used  in
plasma  theory \cite{zhdanov_2002}.
It should also be noted that that the tools to solve multifluid transport
equations numerically are well established
\cite{gallouet_seguin_2004}.

Multifluid models with application to the disperate-mass binary  mixtures
were studied before in the framework of kinetic theory.
The  two-fluid   theory   was   developed   by   Goldman   and   Sirovich
\cite{goldman_sirovich_1967} for the special case of  Maxwell  molecules,
and later attempts were  made  to  extend  it  to  arbitrary  interaction
potentials \cite{mora_fernandez_1987}.
However, quite elegant but rather ad-hoc derivation of  de  la  Mora  and
Fernandez-Feria  \cite{mora_fernandez_1987}  is  restricted   to   binary
mixtures, and, in spite of the  similarity  of  the  resulting  transport
equations with the present work, the use of "unmatched Maxwellians"  that
are not a solution of any equation creates a lot of confusion  throughout
the treatment, especially when dealing with momentum and energy  transfer
between  the components.
The problem was recently approached also by    Kerkhof    and    Geboers
\cite{kerkhof_geboers_2005}.
Relating each component distribution function  to  its  own  hydrodynamic
velocity, they obtained the momentum balance in the form similar  to  the
Maxwell-Stefan equations.
However, in their derivation inaccuracies are present \cite{kerkhof_2007}
and  thus  serious reconsideration of the result is required.
We like to stress that the present derivation  is  not  related  to  ref.
\cite{kerkhof_geboers_2005}, although Kerkhof  has  also  contributed  to
this work.
We  should  also  mention   the   little-known   paper   of   Struminskii
\cite{struminskii_1974} that we came  across  at  the  latest  stages  of
development.
To derive the equations  for  multicomponent  gas  transport  Struminskii
employed the same basic idea of modified scaling  that  is  used  in  the
present work.
However our treatment is very different from given there,  and  we  claim
our results to  be  more  general  and  accurate  compared  to  those  of
Struminskii.
More detailed comparison with this and other papers is given in the  text
where appropriate.

In the present paper we aim to obtain the perturbative  solution  of  the
Chapman-Enskog type for the Boltzmann equation based on extended  set  of
observables $(n_i, \vvec_{i0},  T_i)$,  and  to  establish  equations  of
motion for these variables, relating the transport  coefficients  to  the
microscopic properties of the mixture components.
The  paper  is  organized  as  follows:  In  section  II  we  suggest   a
modification to the  standard  density  scaling  scheme  of  solving  the
Boltzmann equation.
Equations for the distribution functions as  well  as  the  momentum  and
energy balance equations are given  there  for  the  arbitrary  order  in
the formal scaling parameter.
In section III we derive explicit expressions in the 0-th order in formal
scaling parameter, while section IV contains the results  of  the  higher
orders of perturbation theory. 
Discussion of the results is given in section V.

\section{Modified Density Scaling}
We consider the multicomponent Boltzmann equation
\begin{equation}\label{eq:blzmn}
  \mathfrak{D}f_i = \sum_{j}{J\{f_if_j\}},
\end{equation}
which describes the evolution
$\mathfrak{D} f_i = (\frac{\partial}{\partial  t} + \vvec_i\cdot\bnabla_{\bf r} + {\bf F}_i\cdot\bnabla_{\vvec_i})f_i$
of the component $i$ distribution function $f_i$ due to  collisions  with
particles of the same and other components.
The pairwise collisional operator is given by
$J\{f_if_j\} = \int\!d\vvec_j d\bm k_{ij}'\, |\vvec_i - \vvec_j|\sigma_{ij}(|\vvec_i - \vvec_j|,\bm k_{ij}|\bm k_{ij}')({f'}_i{f'}_j - f_if_j)$,
where $\bm k_{ij}$ is the  unit  vector  in  the  direction  of  relative
velocity, and primed quantities denote those after collision.
Here we only consider the case of spherically symmetric interactions, such
that the crossection $\sigma_{ij}(|\vvec_i - \vvec_j|,\bm k_{ij}|\bm k_{ij}')$ 
to change the direction of  relative  velocity  from 
$\bm k_{ij}$ to $\bm k_{ij}'$ depends only on 
$\bm k_{ij}' \cdot \bm k_{ij}$.
We look for the  solution  of  equation  (\ref{eq:blzmn})  based  on  the
set of observables ${\bf \bbeta}_i = (n_i, \vvec_{i0}, T_i)$ including the
hydrodynamic velocities ($\vvec_{i0}$) and temperatures ($T_i$) for  each
of the components.

Following the standard procedure of Enskog we introduce the formally small
scaling  parameter  $\varepsilon$,  put  equal  to  $1$  at  the  end  of
derivation.
The solution is then given in terms  of a series in this parameter
$f_i = \sum_n \varepsilon^n f_i^{(n)}$.
In  the  classical  treatment  this  parameter  is  used  to  scale   the
left-hand part  of  equation  (\ref{eq:blzmn})  which  is  equivalent  to
scaling the density of the mixture.
Such a scaling is very natural  if  one  is  going  to  use  the  set  of
observables including the hydrodynamic velocity and  the  temperature  of
the mixture as a whole $(n_i, \vvec_0, T)$.
Here,  instead,  we  suggest  a  bit  different   scaling   of   equation
(\ref{eq:blzmn})
\begin{equation}\label{eq:newscaling}
  \varepsilon[\mathfrak{D}f_i - \sum_{j \neq i}J\{f_i f_j\}] = J\{f_i f_i\}.
\end{equation}
This corresponds to scaling only the chosen $i$-th component density, and
allows the  zero  order  solutions  to  be  centered  around  different
velocities, which is convenient for deriving  the  species  momentum  and
energy balance equations.
It should be noted that  the  similar  reordering  was  proposed  yet  by
Lorentz \cite{lorentz_1905} for  the  heavy  species  in  disparate  mass
binary mixture, and in modified  form  used  later  by  many  researchers
\cite{grad_1960,chmieleski_ferziger_1967,goebel_johnson_1976}.
However they use such a  reordering  of  the  terms  to  collect  all  the
relevant contributions in equations of the  same  order,  thus  requiring
different  scaling  patterns  for  different   components,   and   as   a
consequence, the different transport equations.
Here instead, we use the  same  form  of  scaling  for  every  component,
employing  it  as  a  formal  tool,  which  allows   to   construct   the
contributions from observable  derivatives  up  to  desired order.
Namely,  in  the  present  paper  we  limit  ourselves  to  the   thermal
conductivity, thermal diffusion and  viscous  contributions.
Using the same form of scaling independently of component  masses  allows
to treat the mixtures with arbitrary number of  components,  but  on  the
other hand it leads to slower convergence of the series in $m_i/\mu_{ij}$
for  some transport coefficients in the case of very light particles.

Expanding  unknown   functions   $f_i$,   operators   $\mathfrak{D}$  and
$J\{f_if_j\}$ as well as the time derivatives of observables
$\partial\bbeta_i/\partial t$ in series in scaling parameter  $\epsilon$,
and assuming after Enskog that the time enters  only  implicitly  through
$\bm \beta$ and the spatial gradients of $\bm \beta$
\begin{equation}
\begin{aligned}
  f_i({\bf r}, \vvec_i, t)
    = f_i({\bf r}, \vvec_i; \bm \bbeta, {\bf\bnabla}_r \bbeta, \ldots)
    = \sum_n \epsilon^n f_i^{(n)},  \\
  \frac{\partial}{\partial t}  \bbeta_i({\bf r}, t)
    \equiv {\bm \Phi}_i({\bf r}; \bbeta, \bnabla_r\bbeta, \ldots)
    = \sum_n \epsilon^n {\bm \Phi}^{(n)}_i.
\end{aligned}
\end{equation}
we arrive at the following equations for the contribution  of  the  order
$n+1$ in a scaling parameter
\begin{equation}\label{eq:fnplus1}
\begin{aligned}
  J\{f_i^{(n+1)}f_i^{(0)}\} +   J\{f_i^{(0)}f_i^{(n+1)}\}
    &= \mathfrak{D}^{(n)}f_i - \sum_{j\neq i}J^{(n)}\{f_if_j\}
      -\sum_{k=1}^{n}J\{f_i^{(k)}f_i^{(n-k+1)}\}, \\
  J\{f_i^{(0)}f_i^{(0)}\} &= 0,
\end{aligned}
\end{equation}
where the following notation is introduced
\begin{equation}
\begin{aligned}
  \mathfrak{D}^{(n)}f_i
   &= \sum_{m=0}^{n}
         \left(
           \bm\Phi^{(m)}\cdot\bnabla_{\bbeta}
          +\bnabla_{\bf r} \bm\Phi^{(m)}:\bnabla_{\nabla_r\bbeta} + \ldots
         \right) f^{(n-m)}
      +(\vvec\cdot\bnabla_r + \bm F \cdot \bnabla_{\vvec})f^{(n)},  \\
  J^{(n)}\{f_if_j\}
   &= \sum_{m=0}^{n} J\{f_i^{(m)}f_j^{(n-m)}\}.
\end{aligned}
\end{equation}
The right-hand side of equations  (\ref{eq:fnplus1})  contains  only  the
functions which are known from the lower order calculations. 
The left-hand side, containing the functions to be  calculated,  includes
only collisional operators for the species of the same type.

The solubility condition for equations (\ref{eq:fnplus1}), demanding that
the right part of  equation  must  be  orthogonal  to  all  solutions  of
the homogeneous equation, read as (the scaling  parameter  $\epsilon$  is
already put equal to $1$ here)
\begin{equation}
  \frac{\partial \bm \beta_i}{\partial t}
  + \sum_n\int\! d^3{\mathit v}_i \bm\psi_i (\vvec_i\cdot\bnabla_r + \bm F\cdot\bnabla_{\mathit v})f^{(n)}
  + \sum_n\sum_{j\neq i}\int\! d^3{\mathit v}_i \bm\psi_i J^{(n)}\{f_if_j\}
  = 0.
\end{equation}
As long as the solution of the  homogeneous  equation  (\ref{eq:fnplus1})
consists of collisional invariants 
$\psi_i = (1, m_i\vvec_i, \frac{1}{2}m_ic_i^2)$, with
$\bm c_i = \vvec_i - \vvec_{i0}$ the velocity related to  coordinate  set
moving  with  the  average  $i$-th  component  velocity,  the  solubility
condition provides us with the corresponding  conservation laws.
After some manipulation this can be rewritten as
\begin{equation}\label{eq:cnsrv_n}
\begin{aligned}
  \frac{1}{\rho_i}\frac{d\rho_i}{dt}
    &= -\bnabla\cdot\vvec_{i0}, \\
  \rho_i \frac{d\vvec_{i0}}{dt}
    &= \rho_i\bm F_i - \sum_n\bnabla \cdot \bm P^{(n)}
     + m_i\sum_{n,j}\int\!d^3c_i \bm c_i J^{(n)}\{f_if_j\}, \\
  \rho_i\frac{d u_i}{dt}
    &= -\bigl(\sum_n \bnabla\cdot\bm q_i^{(n)} + \sum_n \bm P_i^{(n)}:\bnabla\vvec_{i0}\bigr)
     + \frac{m_i}{2}\sum_{n,j}\int\!d^3c_i c_i^2 J^{(n)}\{f_if_j\}.
\end{aligned}
\end{equation}
with partial pressure and the heat flow defined as
\begin{equation}
\begin{aligned}
  \bm P_i^{(n)} &= m_i\int\!d^3c_i \bm c_i \bm c_i f_i^{(n)}, \\
  \bm q_i^{(n)} &= \frac{m_i}{2}\int\!d^3c_i c_i^2 \bm c_i f_i^{(n)}.
\end{aligned}
\end{equation}
Here it should be noted that $\frac{1}{2}m_ic_i^2$ can be considered as a
collisional invariant only in a limited sense as it conserves only  under
collisions of the particles of the same type.
This leads to the sum of the sources
$\sum_{i,j}\int\!d^3c_i \frac{m_i c_i^2}{2}J^{(n)}\{f_if_j\}$ in the heat
balance equation of the set (\ref{eq:cnsrv_n}) being  not equal to zero.
However there is no contradiction here with the  energy  conservation  as
the temperatures for each component are  defined  in  its  own  reference
frame, and thus the energy balance would have also included the terms  of
the $\rho_i v_{i0}^2/2$ type.
Additionally, these source terms can only be calculated exactly  in  some
special cases, and in what follows we expand them assuming  the  velocity
$\vvec_{i0} - \vvec_{j0}$ and temperature $T_i - T_j$  differences  are
small and  keep  only  the  first  nonvanishing  term  of expansion.
Thus, resulting equations should be  understood  as  a  linear  in  above
mentioned differences approximation, while the sum of the source terms in
(\ref{eq:cnsrv_n}) is obviously a contribution of the next order.
Equations  (\ref{eq:cnsrv_n})  can  be  seen  as   the   single-component
balance  equations  with  the  momentum  and  energy  transport  due   to
interaction between the components included as a source terms.
This approach is equivalent to considering the single component system in
a field of hydrodynamic forces exerted by  the  flow  through  the  other
components. 
We like to note however that quantities  $\bm  P_i$  and  $\bm  q_i$  can
depend on the properties of the other components  through  the  functions
$f_i^{(n)}$.

Similarly to original  Enskog  approach  we  derive  here  the  transport
equations in a  general  form  with  all  contributions  determined  from
equations (5,8).
Calculation of these  contributions  is  then  straightforward  (although
elaborate),  with  all assumptions and approximations being very clear.
This should be  contrasted  to  the  previous  derivations,  which  often
require cumbersome argumentation to justify some steps in there.

\section{Zero-order approximation}
The  solution  of  the  equation  (\ref{eq:fnplus1})  in  the  $0$  order
approximation and resulting in  the  proper  observables  values  is  the
Maxwell distribution  related  to  the  frame  moving  with  the  average
velocity  of the $i$-th component $\vvec_{i0}$
\begin{equation}
  f_i^{(0)} = n_i(\gamma_i/2\pi)^{3/2}\exp\left\{-\gamma_i(\vvec_i-\vvec_{i0})^2/2\right\},
\end{equation}
where $n_i$ is the number density of the $i$-th component,
$\gamma_i = m_i/T_i$  with  $m_i$  being  the  mass  of  the  component's
molecules  and  $T_i$  its  kinetic  temperature  (we  prefer  using  the
energetic temperature scale, thus saving on $k_B$).
The resulting partial pressure and the  heat  flow are
\begin{equation}
\begin{aligned}
  \bm P_i^{(0)} &= p_i \bm I = n_iT_i \bm I, \\
  \bm q_i^{(0)} &= 0,
\end{aligned}
\end{equation}
where $\bm I$ is the unit tensor.

The collision integral $J^{(0)}\{f_if_j\}$ can be expanded in  series  in
velocity and temperature difference between the components.
Then, assuming these differences  are  small,  compared  to  the  thermal
velocity and the temperature of the component  correspondingly,  we  keep
only the first nonvanishing terms of expansion.
The  contributions  of  the  zero  order  in  velocity  and   temperature
differences are identically zero, while the  most  general  form  of  the
first order contribution would be
\begin{equation}\label{eq:01b_J0}
  J^{(0)}\{f_if_j\}
    = -f_i^{(0)}Q^{(\vvec)}(c_i)(\vvec_{i0}-\vvec_{j0})\cdot {\bm c}_i
      -f_i^{(0)}Q^{(T)}(c_i)(T_i - T_j).
\end{equation}
Explicit form of the scalar functions $Q^{(\vvec)}(c_i)$ and
$Q^{(T)}(c_i)$ is rather complicated, and for the time being we keep them
undefined.
In the end it turns out that we only need to know  the  even  moments  of
these functions with respect to $c_i$ and not the functions themselves.
Calculation of these moments, using equation (\ref{eq:01b_J0}) gives
\begin{equation}\label{eq:intc2Q}
\begin{aligned}
  \int\!\!d^3\!c_i\, c_i^{2n} f_i^{(0)} Q^{(T)}(c_i)
    &= -n_i
    		\frac{\mu_{ij}}{T_iT_j}
		    \frac{\Gamma(2n+2)}{2^{2n}}
		    \left(\frac{\gamma_{ij}}{2}\right)^{-1}
		    \left(\frac{\gamma_i + \gamma_j}{2}\right)^{-n}
        \sum_{m=1}^{n}\frac{\nu_{ij}^{(1;m)}}{\Gamma(n-m+1)}, \\
  \int\!\!d^3\!c_i\, c_i^{2n+2} f_i^{(0)} Q^{(\vvec)}(c_i)
    &= -n_i
		    \frac{\Gamma(2n+4)}{2^{2n+1}(n+1)}
        \left(\frac{\gamma_i+\gamma_j}{2}\right)^{-n}
        \sum_{m=1}^{n+1}\frac{\nu_{ij}^{(1;m)}}{\Gamma(n-m+2)},
\end{aligned}
\end{equation}
where the hierarchy of frequencies is introduced
\begin{equation}
  \nu_{ij}^{(l;m)}
    = \frac{n_j}{(2\pi\gamma_{ij})^{1/2}}
      \frac{m 2^{2m+3}}{\Gamma(2m+2)}
      \left(\frac{\mu_{ij}}{m_i}\right)^{2m-1}
      \left(\frac{\gamma_i+\gamma_j}{\gamma_{ij}}\right)^{m-1}
      \int_{0}^{\infty}\!\!d\xi\, \xi^{2m+3}e^{-\xi^2}Q_{ij}^{(l)}(g),
\end{equation}
with $\xi=\gamma_{ij}g/2$, $\mu_{ij}$ is the reduced mass of the components
$i$ and $j$, $\gamma_{ij}$ is similarly introduced reduced  quantity, and
$Q_{ij}^{(l)}(g)=\int d\Omega \sigma_{ij}(g,\theta)(1-\cos^l\theta)$ is the
$ij$ transport cross section of the $l$-th order.
$Q_{ij}^{(1)}(g)$  entering  the  zero  order  collision   integral   is
sometimes referred to as a diffusion or momentum transfer cross section.
When all components of the mixture  have  the  same  kinetic  temperature
these frequencies can  be  easily  related  to  familiar  Chapman-Cowling
$\Omega$-integrals: 
$\nu_{ij}^{(l,m)}|_{T_i = T_j} = n_j (m_j/m_i)^m (\mu_{ij}/m_j) \Omega_{ij}^{(l,m)}$.
The finite series for the moments of collision integrals were  calculated
by  Kolodner  \cite{kolodner_1957}  for  the  special  case  of   Maxwell
molecules,    and    were    used     by     Goldman     and     Sirovich
\cite{goldman_sirovich_1967} to develop  the  two-fluid  theory  for  the
mixtures of those molecules.
Here we use the similar formalism to derive the multifluid theory for the
species with arbitrary interaction potential.
This   should   be   contrasted    to    the    work    of    Struminskii
\cite{struminskii_1974}, where similar quantities  were  also  calculated
for the case of Maxwell molecules, and implicit assumption made that  the
same expression can be applied to the species with arbitrary  interaction
potential.

Extracting   the   required   moments   of    $J^{(0)}\{f_if_j\}$    from
(\ref{eq:intc2Q}) we can  write  down  the  transport  equations  in  the
following form
\begin{equation}
\begin{aligned}
  \frac{1}{\rho_i}\frac{d\rho_i}{dt}
    &= -\bnabla\cdot\vvec_{i0}, \\
  \rho_i \frac{d\vvec_{i0}}{dt}
    &= \rho_i\bm F_i - \bnabla p_i
     - n_i\sum_{j}\mu_{ij}\nu_{ij}^{(1;1)}(\vvec_{i0}-\vvec_{j0}), \\
  \rho_i\frac{d u_i}{dt}
    &= -p_i\bnabla\cdot\vvec_{i0}
       -3n_i\sum_{j}\frac{\mu_{ij}}{m_i+m_j}\nu_{ij}^{(1;1)}(T_i - T_j).
\end{aligned}
\end{equation}
The $\nu_{ij}^{(1;1)}$ frequency here coincides with the  effective
collisional frequency of the electron commonly used in the plasma  theory
\cite{zhdanov_2002}.
Thus in the zero order approximation we obtain the  Euler  equations  for
the ideal liquid including the Maxwell-Stefan  terms  for  the  heat  and
momentum diffusion due to interaction between different components.
The  zero  order  results  are  in  total  agreement   with   Struminskii
\cite{struminskii_1974}  and   binary   Mora   \cite{mora_fernandez_1987}
equations, allowing to recover the relationship between diffusion  rates,
obtained by the Enskog method in the first approximation.

\section{First Order Approximation}
The first order contribution to the distribution function is conveniently
put in the form $f_i^{(1)} = f_i^{(0)}\phi_i^{(1)}$, introducing the  new
unknown function $\phi_i^{(1)}$.
Using the results  of  the  previous  section  to  calculate
${\mathfrak    D}^{(0)}f_i$    and    substituting     expression     for
$J^{(0)}\{f_if_j\}$ we can rewrite equation  (\ref{eq:fnplus1})  for  the
unknown function $\phi_i$  as follows
\begin{multline}\label{eq:f0plus1}
  n_i^2 I\{\phi_i^{(1)}\}
    = -f_i^{(0)}
      \biggl[ \gamma_i(\bm c_i \bm c_i - \frac{1}{3}c_i^2\bm I):\bnabla\vvec_{i0}
        +\left(\frac{\gamma_i}{2}c_i^2 - \frac{5}{2}\right)\bm c_i \cdot \bnabla\ln T_i \\
        +\sum_j\left(
                 Q^{(\vvec)}(c_i)-\frac{\mu_{ij}}{T_i}\nu_{ij}^{(1;1)}
               \right) \bm c_i\cdot(\vvec_{i0} - \vvec_{j0})
        +\sum_j\left(
                 Q^{(T)}(c_i)
                +\frac{3}{T_i}\left(\frac{3}{2}-\frac{\gamma_i c_i^2}{2}\right)
                 \frac{\mu_{ij}}{m_i+m_j}\nu_{ij}^{(1;1)}
               \right)(T_i - T_j)
      \biggr],
\end{multline}
where
$I\{\phi\} = n^{-2}\int \!d^3 c_1 d\Omega |\bm c- \bm c_1|f^{(0)}f_1^{(0)}(\phi+\phi_1-\phi'-\phi_1')$
is the linearized collision operator \cite{ferziger_kaper_1972}.
This expression differs from the similar single component equation  only
by the `friction' terms proportional to $(\vvec_{i0} -  \vvec_{j0})$  and
$(T_i -  T_j)$,  therefore  the  corresponding  contributions  should  be
included into  the function $\phi_i^{(1)}$.
The structure of this expression is similar to that used by  Goldman  and
Sirovich \cite{goldman_sirovich_1967}, however in the later work of de la
Mora \cite{mora_fernandez_1987}  the  $Q$-terms  arising  from  collision
integrals were omitted.
Taking into account rotational invariance of collision  operator  we  can
write the function $\phi_i$ in the form
\begin{multline}
  \phi_i
    = -\frac{1}{n_i}A(c_i)\bm c_i\cdot\bnabla\ln T_i
      -\frac{1}{n_i}B(c_i)\left( \bm c_i \bm c_i - \frac{1}{3}c^2 \bm I \right):\bnabla\vvec_{i0}
      \\
      -\frac{1}{n_i}\sum_{j\neq i}D_{ij}(c_i)\bm c_i\cdot(\vvec_{i0} - \vvec_{j0})
      -\frac{1}{n_i}\sum_{j\neq i}E_{ij}(c_i)(T_i - T_j).
\end{multline}
The form of the $\phi_i$ function is similar to that used in the work  of
de la Mora \cite{mora_fernandez_1987}, but differs from the  one  in  the
paper of Struminskii \cite{struminskii_1974}, where the unknown functions
were related not to the driving  forces,  or  independent  parameters  of
solution, but rather chosen  in  accordance  with  tensorial  order  of  
$\bm c_i$ weights.
In our opinion this can not be done if one is looking for solution  based
on extended set of observables.

In order for $\phi_i^{(1)}$ to satisfy  equation  (\ref{eq:f0plus1})  the
following identities should hold
\begin{equation}\label{eq:Iabde}
\begin{aligned}
  n_i I\{A(c_i)\bm c_i\}
    &= f^{(0)}\left(\frac{\gamma_ic_i^2}{2} - \frac{5}{2}\right)\bm c_i, \\
  n_i I\{B(c_i)(\bm c_i \bm c_i - \frac{1}{3}c_i^2\bm I)\}
    &= \gamma_if^{(0)}(\bm c_i \bm c_i - \frac{1}{3}c_i^2\bm I), \\
  n_i I\{D_{ij}(c_i)\bm c_i\}
    &= f^{(0)}\left(Q^{(\vvec)}(c_i) - \frac{\mu_{ij}}{T_i}\nu_{ij}^{(1;1)}\right)\bm c_i, \\
  n_i I\{E_{ij}(c_i)\} &= f^{(0)}
                  \left(
                    Q^{(T)}(c_i)
                    +\frac{3}{T_i}\left(\frac{3}{2}-\frac{\gamma_i c_i^2}{2}\right)
                     \frac{\mu_{ij}}{m_i+m_j}\nu_{ij}^{(1;1)}
                  \right).
\end{aligned}
\end{equation}
Since the  proper  values  of  all  observables  are  obtained  from  the
$0$-order distribution function, several conditions should be applied  to
the function $\phi_i$, such that $f_i^{(1)}$ provides no contribution  to
observables.
In terms of functions $A(c_i), B(c_i)$ and $E(c_i)$ these conditions read
as
\begin{equation}\label{eq:ade_cnstrn}
\begin{aligned}
  \int\!d^3\!c\, c^2 f^{(0)}A(c) &= 0, \\
  \int\!d^3\!c\, c^2 f^{(0)}D_{ij}(c) &= 0, \\
  \int\!d^3\!c\, c^2 f^{(0)}E_{ij}(c) &= \int\!d^3\!c\, f^{(0)}E_{ij}(c) = 0.
\end{aligned}
\end{equation}
which determines the function $\phi_i$ uniquely.

To  determine  the  functions  $A(c),  B(c),  D(c),  E(c)$  we  use   the
variational approach \cite{kohler_1948,ferziger_kaper_1972}, i.e. we look
for the functions that maximize the corresponding bracket  integrals  and
satisfy (\ref{eq:Iabde},\ref{eq:ade_cnstrn}).
Here we should mention that unlike in the classical treatment it  is  not
immidiately obvious that the entropy production would also  be  maximized
by this procedure, and the links  between  the  current  development  and
nonequilibrium thermodynamics require a separate study.
Expanding unknown functions into series we can obtain the simple equations
to determine the expansion coefficients.
The functions $A(c)$ and $B(c)$ here are exactly the same as in classical
single-component approach, thus resulting in the `partial' viscosity  and
thermal conductivity, equal to those the $i$ component would have by  its
own at the same number density $n_i$ and temperature $T_i$ it  has  in  a
mixture.
Expressions for these functions are well known, but we provide them  here
again for completeness. 
Similarly to $A(c)$ and $B(c)$ the functions $D(c)$ and $E(c)$ are  most
easily determined if we expand them  in  series  of  associated  Laguerre
polynomials \cite{kumar_1966,ferziger_kaper_1972}
\begin{equation}\label{eq:abde_exp}
\begin{aligned}
  A_i(c_i) &= -(\gamma_i/2)^{1/2}\sum_{n=1}^{\infty}a_{ij}^{(n)} L_n^{(3/2)}(\gamma_i c_i^2/2), \\
  B_i(c_i) &= (\gamma_i/2)\sum_{n=0}^{\infty}b_{ij}^{(n)} L_n^{(5/2)}(\gamma_i c_i^2/2), \\
  D_{ij}(c_i) &= -(\gamma_i/2)\sum_{n=1}^{\infty}d_{ij}^{(n)} L_n^{(3/2)}(\gamma_i c_i^2/2), \\
  E_{ij}(c_i) &= \sum_{n=2}^{\infty}e_{ij}^{(n)} L_n^{(1/2)}(\gamma_i c_i^2/2).
\end{aligned}
\end{equation}
Some terms in these series are omitted here to satisfy conditions
(\ref{eq:ade_cnstrn}).

Using  the  moments  of  the   the   functions   $Q^{(\vvec)}(c_i)$   and
$Q^{(T)}(c_i)$ (\ref{eq:intc2Q}) we can  write  down  the  simple  matrix
equations to determine  the  expansion  coefficients  $e_{ij}^{(m)}$  and
$d_{ij}^{(m)}$ (the equations for $a_i$ and $b_i$ coefficients are  again
provided here for completeness)
\begin{equation}\label{eq:eijdij}
\begin{aligned}
	\sum_{m=1}^{\infty}
    a_i^{(m)}
    	\left[
    		\left(\frac{\gamma_i}{2}\right)^{1/2} \! \bm c_i 
    		L_{m}^{(3/2)} \! \left(\frac{\gamma_i c_i^2}{2}\right), %
        \left(\frac{\gamma_i}{2}\right)^{1/2}\bm c_i 
    		L_{n}^{(3/2)}\left(\frac{\gamma_i c_i^2}{2}\right)
      \right]
	&= \frac{4}{5}\delta_{n1}, \\
	\sum_{m=0}^{\infty}
    b_i^{(m)}[(\gamma_i/2)(\bm c_i \bm c_i - \frac{1}{3}c_i^2\bm I)L_{m}^{(5/2)}(\gamma_{i}c_{i}^2/2),
              (\gamma_i/2)(\bm c_i \bm c_i - \frac{1}{3}c_i^2\bm I)L_{n}^{(5/2)}(\gamma_{i}c_{i}^2/2)]
  &= \frac{2}{T_i}\delta_{n0}, \\
  \sum_{m=1}^{\infty}
    d_{ij}^{(m)}[(\gamma_i/2)^{1/2}\bm c_iL_{m}^{(3/2)}(\gamma_{i}c_{i}^2/2),
							(\gamma_i/2)^{1/2}\bm c_iL_{n}^{(3/2)}(\gamma_{i}c_{i}^2/2)]
  &= K_{ij}^{(n)}, \\
  \sum_{m=2}^{\infty}
    e_{ij}^{(m)}[L_{m}^{(1/2)}(\gamma_{i}c_{i}^2/2),L_{n}^{(1/2)}(\gamma_{i}c_{i}^2/2)]
  &= \frac{1}{T_i}\frac{\mu_{ij}}{m_j}K_{ij}^{(n-1)}. \\
\end{aligned}
\end{equation}
where the bracket integral is defined as
$[\bm F,\bm G] = Tr\int\!d^3c_i\, \bm G \otimes I\{\bm F\}$  and  the  
shorthand  notation
$K_{ij}^{(n)}
		= \frac{4 \Gamma(n+5/2)}{\sqrt{\pi}}
			\sum_{l=0}^n
        \frac{(-)^l\nu_{ij}^{(1;l+1)}}{\Gamma(n-l+1)}
        \left(1-\frac{\mu_{ij}}{\mu_j}\right)^{n-l}
        \left(\frac{\mu_{ij}}{\mu_j}\right)^l$
is introduced.

Provided equations (\ref{eq:eijdij}) are solved and  thus  the  functions
$A(c)$, $B(c)$, $D(c)$ and $E(c)$ are known, we can calculate  the  first
order contributions to equations (\ref{eq:cnsrv_n}).
For the pressure and the  heat  flow  those read as
\begin{equation}
\begin{aligned}
  \bm P_i^{(1)} &= -b_i^{(0)}T_i\bm S_i, \\
  \bm q_i^{(1)} &= -\frac{5}{4}a_i^{(1)}\bnabla T_i 
  								 -\frac{5T_i}{4}\sum_j d_{ij}^{(1)} (\vvec_{i0} - \vvec_{j0}).
\end{aligned}
\end{equation}
Here, the partial viscosity and heat conductivity are given by the  first
expansion coefficients $b_i^{(0)}T_i/2$ and $5a_i^{(1)}/4$ 
correspondingly.
There is also a term describing the  heat  flow  due  to  interaction  of
the  flows  moving  with  different   velocities,   however   temperature
differences do not contribute to the pressure.
Instead there appears a term in the first order momentum source  integral
having the same effect.
The partial viscosity and heat conductivity here are determined  by  only
the    first    nonvanishing    coefficients    from    the     expansion
(\ref{eq:abde_exp}).
On contrary the contributions from the friction terms, are given in terms
of series including the whole list of expansion coefficients.
Up  to  the  terms  bilinear  in  gradients  of  observables  and   their
differences  the source terms in the first order calculate
\begin{multline}
  m_i\int d^3\!c_i\,{\bm c}_i J^{(1)}\{f_if_j\}
    = -\mu_{ij}^{(\nabla T)}\bnabla T_i - \mu_{ji}^{(\nabla T)}\bnabla T_j
     -(\mu_{ij}^{(T\nabla T)}\bnabla T_i + \mu_{ji}^{(T\nabla T)}\bnabla T_j)(T_i - T_j) \\
     -(\mu_{ij}^{(Sv)}{\bm S}_i + \mu_{ji}^{(Sv)}{\bm S}_j)\cdot(\vvec_{i0}-\vvec_{j0})
     -\sum_k (\mu_{ijk}^{(v)}(\vvec_{i0}-\vvec_{k0}) + \mu_{jik}^{(v)}(\vvec_{j0}-\vvec_{k0})),
\end{multline}
and
\begin{multline}
  \frac{m_i}{2}\int d^3\!c_i\,c_i^2 J^{(1)}\{f_if_j\}
    = -(\lambda_{ij}^{(v\nabla T)}\bnabla T_i 
        +\lambda_{ji}^{(v\nabla T)}\bnabla T_j)\cdot(\vvec_{i0}-\vvec_{j0}) \\
      -\sum_k (\lambda_{ijk}^{(T)}(T_i-T_k) + \lambda_{jik}^{(T)}(T_j-T_k)).
\end{multline}
The transport coefficients  $\mu^{()}$  and  $\lambda^{()}$  include  the
whole  expansions   $\{a^{(n)}\}$,   $\{b^{(n)}\}$,   $\{d^{(n)}\}$   and
$\{e^{(n)}\}$, and are given in appendix A.
The bilinear terms here are not really important and most probably can be
omitted for most of the practical applications.
The most interesting contribution is  probably  the  one  describing  the
thermal diffusion which explicitly appears as a result  of  interactions
between different components in the current treatment.
If the components share the same value of $m_i$ and $T_i$  (that  is  for
example if the single-component flow is virtually divided in two flows of
similar species), this contribution disappears in  total  agreement  with
the classical solution.
Summing the equations to obtain the momentum balance for the mixture as a
whole also eliminates this term.

The other interesting contributions are those with transport coefficients
with $3$ indexes.
Those not only include the first order corrections to the  Maxwell-Stefan
diffusion  terms,  but  also  describe  indirect  interactions   of   two
components through the impact on a third one.
Appearance of such contributions derived  from  the  Boltzmann  equation,
allowing only the pairwise interactions is fascinating but not  totally
unexpected.
Use of the reduced description with some degrees  of  freedom  integrated
out often leads to  many-body effective interactions.
The most famous example of that is probably the interaction of  the  ions
of the same sort in electrolytes.

Thus, in the  first  order  approximation  we  obtain  the  Navier-Stokes
equations which  include  the  corrected  Maxwell-Stefan  terms,  thermal
diffusion  and  also  several  new  terms  bilinear  in   gradients   and
differences of observables.
The partial viscosity and heat conductivity depend only on the properties
of  the  component  $i$,  thus  totally  neglecting  $ij$   correlations.
Also the time evolution of $v_{i0}$ depends on the strain tensor  of  the
component $i$ and does not include the strain of  the  other  components,
which is not physical.
This situation can be understood if we notice  that  the  formal  density
scaling (\ref{eq:newscaling}) we are using, leads to mixing the terms  of
different orders of the scaling parameter $\epsilon$ as compared  to  the
standard treatment, and tends to give more weight to  $ii$  interactions.
Thus,  although,  the  second  order  distribution  function   does   not
contribute to viscous forces in the  classical  approach,  here  we  need
to proceed to higher orders to include the $ij$ viscous interactions.

As many contributions from the  first  order  treatment  can  already  be
considered excessive, we are not going to do the complete analysis of the
second order solution.
Instead we are only interested  in  correcting  the  viscosity  and  heat
conductivity including the terms resulting from $ij$ correlations.
As previously, introducing the  auxiliary  function  $\phi_i^{(2)}$  such
that    $f_i^{(2)}=f_i^{(2)}\phi_i^{(2)}$     we     rewrite     equation
(\ref{eq:fnplus1}) in the form
\begin{equation}
  -n_i^2 I\{\phi_{i}^{(2)}\}
      = \frac{\partial_{1}f_{i}^{(0)}}{\partial t}
       +\frac{\partial_{0}f_{i}^{(1)}}{\partial t}
       +(\bm c\cdot \bm \nabla_r + \bm F\cdot\bm \nabla_v) f_i^{(1)}
       -\sum_{j\neq i}J^{(1)}\{f_i f_j\}
       -J\{f_i^{(1)}f_{i}^{(1)}\}.
\end{equation}
Here $\sum_{j\neq i}J^{(1)}\{f_i f_j\}$  is  the  only  term  containing  the
required contributions, the most general form of which is given by
\begin{equation}
  J^{(1)}\{f_i f_j\} \propto -f_i^{(0)}
    \left[
    \begin{array}{c}
       Q_{ij}^{(S_i)}(c_i)(\bm c_i \bm c_i - \frac{1}{3} c_i^2 \bm I):S_i
      +Q_{ij}^{(\nabla T_i)}(c_i)\bm c_i\cdot \nabla T_i \\
      +Q_{ij}^{(S_j)}(c_i)(\bm c_i \bm c_i - \frac{1}{3} c_i^2 \bm I):S_j
      +Q_{ij}^{(\nabla T_j)}(c_i)\bm c_i\cdot \nabla T_j
    \end{array}
    \right].
\end{equation}
We use the sign $\propto$ to indicate that all  irrelevant  contributions
are omitted here.
The functions $Q_{ij}^{()}(c_i)$ can  be  left  undefined  for  the  time
being, as for the later analysis we will only  need  their  moments  with
respect to even powers of $c_i$ which are easier to  calculate  then  the
functions themselves.

Similarly to the previous section, the viscous contribution to the second
order pressure tensor is found
\begin{equation}
  \bm P_i^{(2)}
    = -T_i \sum_{j\neq i}b_{ij}^{(2;0)} \bm S_i
      -T_i \sum_{j\neq i}b_{ji}^{(2;0)} \bm S_j,
\end{equation}
and nondiagonal part of the heat conductivity reads as
\begin{equation}
  \bm q_i^{(2)} 
  	= -\frac{5}{4} \sum_{j\neq i}a_{ij}^{(2;1)} \bnabla T_i
  		-\frac{5}{4} \sum_{j\neq i}a_{ji}^{(2;1)} \bnabla T_j. 
\end{equation}
The expansion coefficients again can be  found  from  the  simple  matrix
equations
\begin{equation}\label{eq:2rdrba}
\begin{aligned}
  \sum_{m=0}^{\infty}
    b_{ij}^{(2;m)}[(\bm c_i \bm c_i - \frac{1}{3} c_i^2 \bm I) L_m^{(5/2)}(\gamma_ic_i^2/2),
                   (\bm c_i \bm c_i - \frac{1}{3} c_i^2 \bm I) L_n^{(5/2)}(\gamma_ic_i^2/2)]
   &= K_{ij}^{(2b;n)}, \\
  \sum_{m=1}^{\infty}
    a_{ij}^{(2;m)}[(\gamma_i/2)^{1/2} \bm c_i L_m^{(3/2)}(\gamma_ic_i^2/2),
    							 (\gamma_i/2)^{1/2} \bm c_i L_n^{(3/2)}(\gamma_ic_i^2/2)]
   &= K_{ij}^{(2a;n)}
\end{aligned}
\end{equation}
Expressions for the coefficients $K_{ij}^{(2;n)}$ are given in  Appendix
B.
When the mixture is treated as a  whole  and  thus  all  components  have
approximately the same  rate  of  shear  and  temperature  gradient,  the
viscosity and heat conductivity for the mixture can be found simply as  a
sum  of  all  the components of corresponding matrices.
\section{Discussion}
We investigate the perturbative solution to the Boltzmann  multicomponent
kinetic equation based on the set of observables including each component
velocity and the temperature.
The corresponding set of the species balance equations is derived.
The new solution is obtained through modification of the  formal  density
scaling scheme of Enskog, such that the  density  of  each  component  is
scaled independently.
This approach allows for succesive development,  with  clear  assumptions
and approximations, which should be contrasted to rather ad-hoc  previous
attacks on the problem.
Omittance or addition of the terms  in  some  intermediate  equations  as
compared to  the  similar  works,  as  discussed  in  the  text,  follows
directly from the scaling equation (\ref{eq:newscaling}).

We show that  the  distribution  functions  related  to  each  component
velocity  appear  naturally  using  this  approach,  and  the  solubility
conditions immediately provide us with the balance equations for species.
The zero order equations are  the  Euler  equations  for  each  component
including the heat and momentum transport between components through  the
familiar  Maxwell-Stefan  diffusion  term.
As a next  approximation  we  obtain  the  Navier-Stokes  equations  with
partial viscosities equal to those the component would have in absence of
the other.
Apart from that there are viscous contributions from the shear  of  other
components.
The interaction between  components  leads  to  appearance  of  the  heat
diffusion and also some  corrections  to  the  Maxwell-Stefan  diffusion
term.
These corrections include apart from direct interactions between the  two
components indirect ones, through perturbing the other components flows.
The heat balance equations posses the similar properties.
To  summarize  we  provide  here  the  final  equations  describing   the
multicomponent flow
\begin{equation}
\begin{aligned}
  \frac{1}{\rho_i}\frac{d\rho_i}{dt}
    =& -\bnabla\cdot\vvec_{i0}, \\
  \rho_i \frac{d\vvec_{i0}}{dt}
    =& \rho_i\bm F_i - \bnabla p_i
     + 2\eta_{i}\bnabla \bm S_i
     + 2\sum_{j\neq i}(\eta_{ij}\bnabla\bm S_i + \eta_{ji}\bnabla\bm S_j) 
     -\sum_{j\neq i}(\mu_{ij}^{(\nabla T)}\bnabla T_i + \mu_{ji}^{(\nabla T)}\bnabla T_j) \\ 
    &- n_i\sum_{j}\mu_{ij}\nu_{ij}^{(1;1)}(\vvec_{i0}-\vvec_{j0})
	   -\sum_k (\mu_{ijk}^{(v)}(\vvec_{i0}-\vvec_{k0}) + \mu_{jik}^{(v)}(\vvec_{j0}-\vvec_{k0})), \\
  \rho_i\frac{d u_i}{dt}
    =& -p_i\bnabla\cdot\vvec_{i0} 
       +\bnabla \lambda_i\bnabla T_i 
       + \sum_{j\neq i}(\bnabla \lambda_{ij}\bnabla T_i + \bnabla \lambda_{ji}\bnabla T_i) \\
      &+2\eta_i \bm S_i : \bnabla\vvec_{i0} 
       +2\sum_{j\neq i}(\eta_{ij}\bm S_i + \eta_{ji}\bm S_j):\bnabla\vvec_{i0} \\
      &-3n_i\sum_{j}\frac{\mu_{ij}}{m_i+m_j}\nu_{ij}^{(1;1)}(T_i - T_j)
       -\sum_k (\lambda_{ijk}^{(T)}(T_i-T_k) + \lambda_{jik}^{(T)}(T_j-T_k)),
\end{aligned}
\end{equation}
where the transport coefficients are deifned  in  the  text  through  the
propeties of the compoenents.

The convergence of the scheme is not questioned in the paper, however one
may argue that apart from using an extended set of observables the  whole
procedure can be seen as mixing the terms of different  orders  (compared
to the classical treatment),  and  thus  convergence  of  both  solutions
should be about the same, suggesting the range of Knudsen  numbers  where
it is applicable.
Another issue is the convergence  of  the  transport  coefficients  which
depend on the whole first  order  distribution  function,  and  not  just
include the first expansion coefficient.
As we see, convergence of these series depend strongly on  the  value  of
$0 < \gamma_{ij}/\gamma_j  <  1$. 
In cases when the temperatures of the different components are about  the
same, and the impact of the temperature  gradient  is  more  significant,
which is true for many chemical engineering applications, the above ratio
can  be  replaced  by  $\mu_{ij}/m_j$,  thus  removing  the   complicated
temperature dependency from the transport coefficients. 
In the case when the gradients are small, but the temperatures themselves
are  very  different,  as  it  may  happen   for   example   in   plasma,
the $\gamma_{ij}/\gamma_j$ ratio should be kept,  but  it  varies  slowly
again.
The  only  potential  problems  may  appear  when  both   gradients   and
temperature  differences  are  big   enough,   leading   to   complicated
temperature dependency of the transport coefficients, however  this  case
is  clearly  out  of  scope  of  the  present  theory.
The present study can also be seen as a basis for  the  multifluid  model
used to describe kinetic processes in plasma.
The other potential applications of the  method  presented  here  include
reactive flows and the boundary layer problem for  mixtures  of  rarified
gases, that still generates a lot of discussion in  the  literature  (see
for example \cite{kremer_soares_2006,sharipov_kolempa_2003}).
Some authors also tend to employ the multicomponent  transport  equations
to   describe   the   shock-waves   or    even    turbulence    phenomena
\cite{struminskii_1996}.

Apart from modified scaling, the presented solution follows  closely  the
original Enskog scheme, with which it has more in common then  any  other
work on multicomponent transport cited here.
However some features appear to be very different.
The question of entropy production which is not  obviously  maximized  by
the present solution requires separate study to establish the links  with
the nonequilibrium thermodynamics.
The fact that there are no $ij$  collisions  in  the  left  part  of  the
equation  (\ref{eq:fnplus1})  eliminates  the  need   of   $ij$   bracket
integrals.
In some sense they are replaced by the  set  of  collisional  frequencies
$\nu_{ij}^{(m;n)}$,    closely    related    to    the     Chapman-Enskog
$\Omega$-integrals. 
The  $\nu_{ij}^{(1;1)}$  is  recognized  as  an   effective   collisional
frequency of  electrons  from  the  plasma  theory \cite{zhdanov_2002}. 
Unlike in the classical  solution  the  first  order  solution  does  not
provide us with all the relevant contributions, and the contributions  of
the same nature, such as viscous, are divided between the first  and  the
second order contributions.
This is the consequence of the suggested scaling, which gives more weight
to collisions of the species of the same type.
On the other hand the $ii$ and $ij$ contributions to  the  viscosity  and
heat conductivity are explicitly separated, which makes it convenient to
compare the present results with the Green-Kubo theory for the  transport
coefficients.
It should also be noted that for the special case of binary mixtures  the
form of the transport equations derived here coincide with those obtained
by de la Mora \cite{mora_fernandez_1987}.
The structure of the equations for the transport coefficients however  is
somewhat different, although we do not expect large discrepancies in  the
numerical values.
The  detailed  numerical  study   revealing   importance   of   different
contributions, and  containing  comparison  with  counterflow  and  other
experiments (simulations) is the topic of the future research.

\section*{Acknowledgements}
We like to thank Dick Bedeaux and Joachim Gross for fruitful  discussion.
This work was supported by an ECHO grant $700.540.41$,  from  Netherlands
Organisation for Scientific Research.

\appendix
\section*{Appendix A}
Below we provide the expressions for the transport coefficients  entering
the first order equations of motion.
Lower signs correspond to switching $i$ and $j$ indexes, and $F(a,b;c;x)$
is the hypergeometric function.
\begin{eqnarray*}
  \mu_{ij}^{(\nabla T)} &=& 
    \mp\frac{4}{3\pi^{1/2}}
	     \left(\frac{\gamma_i}{2}\right)^{1/2}
       \sum_{n}a_{i}^{(n)}\Gamma(n+5/2)\sum_{s=0}^{n} (-)^{s}
       	 \left(\frac{\gamma_{ij}}{\gamma_j}\right)^{s}\\
       &&\times 
         \left(
           \frac{F(s+3/2,s-n;s+5/2;\gamma_{ij}/\gamma_j)}{\Gamma(n-s+1)}
          -\frac{\gamma_{ij}}{\gamma_j}
         	 \frac{F(s+5/2,s-n+1;r+7/2;\gamma_{ij}/\gamma_j)}{\Gamma(n-s)(s+5/2)}
         \right) \nu_{ij}^{(1;s+1)}
\end{eqnarray*}
\begin{eqnarray*}
  \mu_{ij}^{(T\nabla T)} &=& 
  	\frac{2}{3\pi^{1/2}}\left(\frac{\gamma_i}{2}\right)^{1/2}
  	\frac{\gamma_i}{T_j}
    \sum_{n}a_{i}^{(n)}\Gamma(n+5/2)\sum_{s=0}^{n}(-)^{s}
    	\left(\frac{\gamma_{ij}}{\gamma_j}\right)^{s}\\
    &&\times 
    \left( 
    \begin{array}{c}
      -\left( 
         \frac{F(s+3/2;s-n;s+5/2;\gamma_{ij}/\gamma_j)}{\Gamma(n-s+1)}
        -\frac{\gamma_{ij}}{\gamma_j}
         \frac{(2s+1)F(s+5/2;s-n+1;s+7/2;\gamma_{ij}/\gamma_j)}{(s+5/2)\Gamma(n-s)}
       \right)\nu_{ij}^{(1;s+1)} \\ 
      +\frac{\gamma_{ij}}{\gamma_j}
       \frac{(s+1)F(s+5/2,s-n+1;r+7/2;\gamma_{ij}/\gamma_j)}{\Gamma(n-s)}
       \nu_{ij}^{(1;s+2)}
    \end{array}
    \right)	
\end{eqnarray*}
\begin{eqnarray*}
\mu_{ij}^{(Sv)} &=&
	\frac{8 \mu_{ij}}{15 \pi^{1/2}}
	\sum_{n}b_{ij}^{(n)}\Gamma(n+7/2)
	\sum_{s=0}^{n}(-)^{s}\left(\frac{\gamma_{ij}}{\gamma_j}\right)^{s} \\
	&&\times
	\left( 
    \begin{array}{c}
      -\frac{m_i}{m_j}
       \sum_{s=0}^{n}
       	 \frac{s F(s+5/2,s-n+1;s+9/2;\gamma_{ij}/\gamma_j)}
       	 			{(s+5/2)( s+7/2) \Gamma(n-s)}
       	 \frac{\gamma_{ij}}{\gamma_j}\nu_{ij}^{(1;s+1)} \\ 
      +\frac{m_i}{m_j}
       \sum_{s=0}^{n}
       \left(
         \frac{4(s+1)F(s+3/2,s-n;s+7/2;\gamma_{ij}/\gamma_j)}
         			{\Gamma(n-s+1)}
        -\frac{\gamma_{ij}}{\gamma_j}
         \frac{16(s+1)F(s+5/2,s-n+1;s+9/2;\gamma_{ij}/\gamma_j)}
        			{(2s+7)\Gamma(n-s)}
       \right) \nu_{ij}^{(1;s+2)} \\ 
      +\frac{\gamma_{ij}}{\gamma_j}
       \sum_{s=0}^{n}
       \frac{2(s+1)}{(2s+7)(2s+9)}
       \left(
       	 \frac{F(s+5/2,s-n+1;s+9/2;\gamma_{ij}/\gamma_j)}{(2s+5)\Gamma(n-s)}
       	-\frac{\gamma_{ij}}{\gamma_j}
       	 \frac{2F(s+7/2,s-n+2;s+11/2;\gamma_{ij}/\gamma_j)}
       	 			{\Gamma(n-s-1)}
       \right)\nu_{ij}^{(1;s+1)} \\
		  +\frac{\gamma_{ij}}{\gamma_j} 
       \sum_{s=0}^{n}
         \frac{4(s+1)F( s+5/2,-n+s+1;s+9/2;\gamma_{ij}/\gamma_j)}
         			{(s+7/2)\Gamma(n-s)}
         \nu_{ij}^{(1;s+2)}	\\
      -\sum_{s=0}^{n}
       \left(
         \frac{5F(s+3/2,s-n;s+7/2;\gamma_{ij}/\gamma_j)}{(2s+5)\Gamma(n-s+1)}
        -\frac{\gamma_{ij}}{\gamma_j}
         \frac{2(4s+5)F(s+5/2,-n+s+1;s+9/2;\gamma_{ij}/\gamma_j)}
         			{(2s+5)(2s+7)\Gamma(n-s)}
       \right) \nu_{ij}^{(1;s+1)}
    \end{array}%
	\right)
\end{eqnarray*}
\begin{eqnarray*}
  \mu_{ijk}^{(v)} &=& 
    \mp\frac{4}{3\pi^{1/2}}
	     \frac{\gamma_i}{2}
       \sum_{n}d_{ik}^{(n)}\Gamma(n+5/2)\sum_{s=0}^{n} (-)^{s}
       	 \left(\frac{\gamma_{ij}}{\gamma_j}\right)^{s}\\
       &&\times 
         \left(
           \frac{F(s+3/2;s-n;s+5/2;\gamma_{ij}/\gamma_j)}{\Gamma(n-s+1)}
          -\frac{\gamma_{ij}}{\gamma_j}
         	 \frac{F(s+5/2,s-n+1;r+7/2;\gamma_{ij}/\gamma_j)}{\Gamma(n-s)(s+5/2)}
         \right) \nu_{ij}^{(1;s+1)}
\end{eqnarray*}
\begin{eqnarray*}
 	\lambda_{ij}^{(v\nabla T)} &=& 
 		\frac{4}{3\pi^{1/2}}\left(\frac{\gamma_i}{2}\right)^{1/2}
 		\sum_{n}a_{i}^{(n)}\Gamma(n+5/2)
    \sum_{s=0}^{n}(-)^{s}\left(\frac{\gamma_{ij}}{\gamma_j}\right)^{s} \\ &&
    \times 
    \left(
    \begin{array}{c}
      \left(\frac{\mu_{ij}}{m_i} - \frac{\gamma_{ij}}{\gamma_j}\right)
      \left( 
        \frac{F(s+3/2,s-n;s+5/2;\gamma_{ij}/\gamma_j)}{\Gamma(n-s+1)} 
       -\frac{\gamma_{ij}}{\gamma_j}
       	\frac{(2s+1)F(s+5/2,s-n+1;s+7/2;\gamma_{ij}/\gamma_j)}
       			 {(s+5/2)\Gamma(n-s)}
      \right) \nu _{ij}^{(1;s+1)} \\ 
     \mp
      \frac{\gamma_{ij}}{\gamma_j}
      \left(
      	\frac{5F(s+5/2,s-n+1;s+7/2;\gamma_{ij}/\gamma_j)}
      			 {(s+5/2) \Gamma(n-s)}
       -\frac{\gamma_{ij}}{\gamma_j}
       	\frac{4F(s+7/2,s-n+2;s+9/2;\gamma_{ij}/\gamma_j)}
       	{(s+7/2)\Gamma(n-s-1)}
      \right)\nu_{ij}^{(1;s+1)}	 \\ 
     \pm
      \frac{\gamma_{ij}}{\gamma_j}
      \frac{2(s+1) F(s+5/2,s-n+1;s+7/2;\gamma_{ij}/\gamma_j)}
      		 {\Gamma(n-s)}
      \nu_{ij}^{(1;s+2)}
    \end{array}
		\right) 
\end{eqnarray*}
\begin{eqnarray*}
\lambda_{ijk}^{(T)} &=&
  \mp\frac{4 T_i}{\pi^{1/2}}
  \sum_{n}e_{ik}^{(n)}\sum_{s=0}^{n-1}
     \frac{(-)^s\Gamma(n+3/2)}{\Gamma(n-a)}
     \left(1-\frac{\gamma_{ij}}{\gamma_j}\right)^{n-s-1}
     \left(\frac{\gamma_{ij}}{\gamma_j}\right)^{s+1}\nu_{ij}^{(1;s+1)}.
\end{eqnarray*}

\appendix
\section*{Appendix B}
Here we provide the expressions for coefficients appearing in  the  right
hand side of equations (\ref{eq:2rdrba}) which determine the  nondiagonal
(in component indexes) viscosity and heat conductivity.  Here  again  the
$ij$ coefficients are obtained by simultaneous replacing all $i$  by  $j$
and contrary, no sign  changes  required  here.  These  expressions  are
rather awkward, and in principle  several  sums  should  be  possible  to
calculate analytically, however  we  failed  to  further  simplify  these
expressions, which anyway would not make significant difference  for  the
numerical calculations of the coefficients.
\begin{eqnarray*}
  K_{ij}^{(2b;n)} &=&
  	-\frac{1}{n_{i}}\frac{\gamma_{i}}{2}\frac{1}{2 \pi ^{1/2}}%
     \sum_{m}b_{i}^{(m)}
     \sum_{l=0}^{m}\sum_{p=0}^{n}
       \frac{(-)^{l+p}}{l!p!}\binom{m+5/2}{m-l}\binom{n+5/2}{n-p}
       \left(\frac{\gamma_{ij}}{\gamma_j}\right)^{l+p}
       \sum_{k=0}^{l}\sum_{q=0}^{p}\binom{l}{k}\binom{p}{q}  \\ &&
     \times 
     \sum_{r=0}^{l+p-k-q}\binom{l+p-k-q}{r}
     \left( 
     \begin{array}{c}
      \frac{W_{l+p,k+q,r}^{(5/2)}}{k+q+1}
       \left(
	      -\frac{1}{3}\nu_{ij}^{(k;(k+q)/2+r)}+2\nu_{ij}^{(k+1;(k+q)/2+r)}
	      -\frac{1}{3}\nu_{ij}^{(k+2;(k+q)/2+r)}
       \right)  \\ 
      +\frac{W_{l+p,k+q,r}^{(5/2)}}{k+q+3}
       \left(
       	 \nu_{ij}^{(k;(k+q)/2+r)}+\frac{2}{3}\nu_{ij}^{(k+1;(k+q)/2+r)}
       	+\nu_{ij}^{(k+2;(k+q)/2+r)}
       \right)  \\
      +\frac{2}{3}
       \frac{W_{l+p,k+q,r-1}^{(5/2)}}{k+q+1}
       \nu_{ij}^{(k;(k+q)/2+r-1)} \\ 
      +\frac{2}{3}
       \frac{W_{l+p,k+q,r+1}^{(5/2)}}{k+q+1}
       \nu_{ij}^{(k+2;k/2+q/2+r+1)} \\ 
      +\frac{4}{3}
       \frac{W_{l+p,k+q,r-1/2}^{(5/2)}}{k+q+2}
       \left(
          \nu _{ij}^{(k;(k+q-1)/2+r)}+\nu_{ij}^{(k+1;(k+q-1)/2+r)}
       \right)  \\ 
      +\frac{4}{3}
       \frac{W_{l+p,k+q,r+1/2}^{(5/2)}}{k+q+2}
       \left( 
       	 \nu_{ij}^{(k+2;(k+q+1)/2+r)}+\nu_{ij}^{(k+1;(k+q+1)/2+r)}
       \right) 
     \end{array}%
     \right)
\end{eqnarray*}
\begin{eqnarray*}
	 K_{ij}^{(2a;n)} &=&
	   \frac{1}{n_{i}}\frac{1}{2\pi^{1/2}}
	   \sum_{m}a_{i}^{(m)}\sum_{l=0}^{m}
	   \frac{(-)^{l+p}}{l!p!}\binom{m+3/2}{m-l}\binom{n+3/2}{n-p}
	   \left(\frac{\gamma_{ij}}{\gamma_j}\right)^{l+p}
     \sum_{k=0}^{l}\sum_{q=0}^{p}\binom{l}{k}\binom{p}{q}	 \\ &&
	  \times 
     \sum_{r=0}^{p+l-k-q}\binom{p+l-k-q}{r}
     \left( 
     \begin{array}{c}
       \frac{W_{l+p,k+q,r-1}^{(3/2)}}{k+q+1}
       \nu_{ij}^{(k;(k+q)/2+r-1)} \\ 
      +\frac{W_{l+p,k+q,r}^{(3/2)}}{k+q+1}
       \nu_{ij}^{(k+1;(k+q)/2+r)} \\ 
      +\frac{W_{l+p,k+q,r-1/2}^{(3/2)}}{k+q+2}
       \left(\nu_{ij}^{(k;(k+q-1)/2+r)}+\nu_{ij}^{(k+1;(k+q-1)/2+r)}\right)
     \end{array}
     \right)
\end{eqnarray*}
Here we have introduced an auxiliary function 
$W_{l,m,n}^{a} = \frac{2^{-2n}\Gamma(l-m/2+n+a)\Gamma(m+2n+4)}{(m+r+1)}$,
and it is also  implied  that  only  contributions  with  integer  upper
indexes  of $\nu_{ij}^{(l;m)}$ enter the summation.

\bibliography{literature}

\begin{thebibliography}{34}
\expandafter\ifx\csname natexlab\endcsname\relax\def\natexlab#1{#1}\fi
\expandafter\ifx\csname bibnamefont\endcsname\relax
  \def\bibnamefont#1{#1}\fi
\expandafter\ifx\csname bibfnamefont\endcsname\relax
  \def\bibfnamefont#1{#1}\fi
\expandafter\ifx\csname citenamefont\endcsname\relax
  \def\citenamefont#1{#1}\fi
\expandafter\ifx\csname url\endcsname\relax
  \def\url#1{\texttt{#1}}\fi
\expandafter\ifx\csname urlprefix\endcsname\relax\def\urlprefix{URL }\fi
\providecommand{\bibinfo}[2]{#2}
\providecommand{\eprint}[2][]{\url{#2}}

\bibitem[{\citenamefont{Chapman}(1917)}]{chapman_1917}
\bibinfo{author}{\bibfnamefont{S.}~\bibnamefont{Chapman}},
  \bibinfo{journal}{Phil.~Trans.~Roy.~Soc.~London}
  \textbf{\bibinfo{volume}{217}}, \bibinfo{pages}{118} (\bibinfo{year}{1917}).

\bibitem[{\citenamefont{Enskog}(1917)}]{enskog_1917}
\bibinfo{author}{\bibfnamefont{D.}~\bibnamefont{Enskog}},
  \emph{\bibinfo{title}{Kinetische Theorie der Vorgange in massig verdunnten
  Gases}} (\bibinfo{publisher}{Diss., Uppsala}, \bibinfo{year}{1917}).

\bibitem[{\citenamefont{Ferziger and Kaper}(1972)}]{ferziger_kaper_1972}
\bibinfo{author}{\bibfnamefont{J.}~\bibnamefont{Ferziger}} \bibnamefont{and}
  \bibinfo{author}{\bibfnamefont{H.}~\bibnamefont{Kaper}},
  \emph{\bibinfo{title}{Mathematical theory of transport processes in gases}}
  (\bibinfo{publisher}{Amsterdam:North-Holland}, \bibinfo{year}{1972}).

\bibitem[{\citenamefont{Grad}(1963)}]{grad_1963}
\bibinfo{author}{\bibfnamefont{H.}~\bibnamefont{Grad}}, \bibinfo{journal}{The
  Physics of Fluids} \textbf{\bibinfo{volume}{6(2)}}, \bibinfo{pages}{147}
  (\bibinfo{year}{1963}).

\bibitem[{\citenamefont{Alavi and Snider}(1998)}]{alavi_snider_1998}
\bibinfo{author}{\bibfnamefont{S.}~\bibnamefont{Alavi}} \bibnamefont{and}
  \bibinfo{author}{\bibfnamefont{R.~F.} \bibnamefont{Snider}},
  \bibinfo{journal}{J.~Chem.~Phys.} \textbf{\bibinfo{volume}{109(9)}},
  \bibinfo{pages}{3452} (\bibinfo{year}{1998}).

\bibitem[{\citenamefont{Chen et~al.}(2001)\citenamefont{Chen, Rao, and
  Spiegel}}]{chen_spiegel_2001}
\bibinfo{author}{\bibfnamefont{X.}~\bibnamefont{Chen}},
  \bibinfo{author}{\bibfnamefont{H.}~\bibnamefont{Rao}}, \bibnamefont{and}
  \bibinfo{author}{\bibfnamefont{E.}~\bibnamefont{Spiegel}},
  \bibinfo{journal}{Phys.~Rev.~E} \textbf{\bibinfo{volume}{64}},
  \bibinfo{pages}{046308} (\bibinfo{year}{2001}).

\bibitem[{\citenamefont{Kerkhof and
  Geboers}(2005{\natexlab{a}})}]{kerkhof_geboers_2005}
\bibinfo{author}{\bibfnamefont{P.}~\bibnamefont{Kerkhof}} \bibnamefont{and}
  \bibinfo{author}{\bibfnamefont{M.}~\bibnamefont{Geboers}},
  \bibinfo{journal}{A.I.Ch.E.~Journal} \textbf{\bibinfo{volume}{51}},
  \bibinfo{pages}{79} (\bibinfo{year}{2005}{\natexlab{a}}).

\bibitem[{\citenamefont{Bearman and Kirkwood}(1958)}]{bearman_kirkwood_1958}
\bibinfo{author}{\bibfnamefont{R.}~\bibnamefont{Bearman}} \bibnamefont{and}
  \bibinfo{author}{\bibfnamefont{J.}~\bibnamefont{Kirkwood}},
  \bibinfo{journal}{J.~Chem.~Phys.} \textbf{\bibinfo{volume}{28}},
  \bibinfo{pages}{136} (\bibinfo{year}{1958}).

\bibitem[{\citenamefont{Snell et~al.}(1967)\citenamefont{Snell, Aranow, and
  Spangler}}]{snell_aranow_1967}
\bibinfo{author}{\bibfnamefont{F.}~\bibnamefont{Snell}},
  \bibinfo{author}{\bibfnamefont{R.}~\bibnamefont{Aranow}}, \bibnamefont{and}
  \bibinfo{author}{\bibfnamefont{R.}~\bibnamefont{Spangler}},
  \bibinfo{journal}{J.~Chem.~Phys.} \textbf{\bibinfo{volume}{47}},
  \bibinfo{pages}{4959} (\bibinfo{year}{1967}).

\bibitem[{\citenamefont{Krishna and
  Wesselingh}(1997)}]{krishna_wesselingh_1997}
\bibinfo{author}{\bibfnamefont{R.}~\bibnamefont{Krishna}} \bibnamefont{and}
  \bibinfo{author}{\bibfnamefont{J.~A.} \bibnamefont{Wesselingh}},
  \bibinfo{journal}{Chemical Engineering Science}
  \textbf{\bibinfo{volume}{52(6)}}, \bibinfo{pages}{861}
  (\bibinfo{year}{1997}).

\bibitem[{\citenamefont{Felderhof}(2003)}]{felderhof_2003}
\bibinfo{author}{\bibfnamefont{B.~U.} \bibnamefont{Felderhof}},
  \bibinfo{journal}{J.~Chem.~Phys.} \textbf{\bibinfo{volume}{118(24)}},
  \bibinfo{pages}{11326} (\bibinfo{year}{2003}).

\bibitem[{\citenamefont{Runstedler}(2006)}]{runstedler_2006}
\bibinfo{author}{\bibfnamefont{A.}~\bibnamefont{Runstedler}},
  \bibinfo{journal}{Chemical Engineering Science}
  \textbf{\bibinfo{volume}{61}}, \bibinfo{pages}{5021} (\bibinfo{year}{2006}).

\bibitem[{\citenamefont{Schimpf and Semenov}(2004)}]{schimpf_semenov_2004}
\bibinfo{author}{\bibfnamefont{M.~E.} \bibnamefont{Schimpf}} \bibnamefont{and}
  \bibinfo{author}{\bibfnamefont{S.~N.} \bibnamefont{Semenov}},
  \bibinfo{journal}{Phys.~Rev.~E} \textbf{\bibinfo{volume}{70}},
  \bibinfo{pages}{031202} (\bibinfo{year}{2004}).

\bibitem[{\citenamefont{Kerkhof and
  Geboers}(2005{\natexlab{b}})}]{kerkhof_geboers_2005(1)}
\bibinfo{author}{\bibfnamefont{P.}~\bibnamefont{Kerkhof}} \bibnamefont{and}
  \bibinfo{author}{\bibfnamefont{M.}~\bibnamefont{Geboers}},
  \bibinfo{journal}{Chemical Engineering Science}
  \textbf{\bibinfo{volume}{60}}, \bibinfo{pages}{3129}
  (\bibinfo{year}{2005}{\natexlab{b}}).

\bibitem[{\citenamefont{Young and Todd}(2005)}]{young_todd_2005}
\bibinfo{author}{\bibfnamefont{J.~B.} \bibnamefont{Young}} \bibnamefont{and}
  \bibinfo{author}{\bibfnamefont{B.}~\bibnamefont{Todd}},
  \bibinfo{journal}{International Journal of Heat and Mass Transfer}
  \textbf{\bibinfo{volume}{48}}, \bibinfo{pages}{5338} (\bibinfo{year}{2005}).

\bibitem[{\citenamefont{Remick and Geankoplis}(1974)}]{remick_geankoplis_1974}
\bibinfo{author}{\bibfnamefont{R.}~\bibnamefont{Remick}} \bibnamefont{and}
  \bibinfo{author}{\bibfnamefont{C.}~\bibnamefont{Geankoplis}},
  \bibinfo{journal}{Chemical~Engineering~Science}
  \textbf{\bibinfo{volume}{29}}, \bibinfo{pages}{121} (\bibinfo{year}{1974}).

\bibitem[{\citenamefont{Stefan}(1871)}]{stefan_1871}
\bibinfo{author}{\bibfnamefont{J.}~\bibnamefont{Stefan}},
  \bibinfo{journal}{Sitzungsber.Osterr.Akademie der Wissensch}
  \textbf{\bibinfo{volume}{63}}, \bibinfo{pages}{63} (\bibinfo{year}{1871}).

\bibitem[{\citenamefont{Maxwell}(1860)}]{maxwell_1860}
\bibinfo{author}{\bibfnamefont{J.~C.} \bibnamefont{Maxwell}},
  \bibinfo{journal}{Philosophical Transactions of the Royal Society}
  \textbf{\bibinfo{volume}{157}} (\bibinfo{year}{1860}).

\bibitem[{\citenamefont{Zhdanov}(2002)}]{zhdanov_2002}
\bibinfo{author}{\bibfnamefont{V.}~\bibnamefont{Zhdanov}},
  \emph{\bibinfo{title}{transport processes in multicomponent plasma}}
  (\bibinfo{publisher}{Taylor \& Francis}, \bibinfo{year}{2002}).

\bibitem[{\citenamefont{Gallouet et~al.}(2004)\citenamefont{Gallouet, Harrard,
  and Seguin}}]{gallouet_seguin_2004}
\bibinfo{author}{\bibfnamefont{T.}~\bibnamefont{Gallouet}},
  \bibinfo{author}{\bibfnamefont{J.-M.} \bibnamefont{Harrard}},
  \bibnamefont{and} \bibinfo{author}{\bibfnamefont{N.}~\bibnamefont{Seguin}},
  \bibinfo{journal}{Mathematical Models and Methods in Applied Sciences}
  \textbf{\bibinfo{volume}{14(5)}}, \bibinfo{pages}{663}
  (\bibinfo{year}{2004}).

\bibitem[{\citenamefont{Goldman and Sirovich}(1967)}]{goldman_sirovich_1967}
\bibinfo{author}{\bibfnamefont{E.}~\bibnamefont{Goldman}} \bibnamefont{and}
  \bibinfo{author}{\bibfnamefont{L.}~\bibnamefont{Sirovich}},
  \bibinfo{journal}{The Physics of Fluids} \textbf{\bibinfo{volume}{10}},
  \bibinfo{pages}{1928} (\bibinfo{year}{1967}).

\bibitem[{\citenamefont{de~la Mora and
  Fernandez-Feria}(1987)}]{mora_fernandez_1987}
\bibinfo{author}{\bibfnamefont{J.~F.} \bibnamefont{de~la Mora}}
  \bibnamefont{and}
  \bibinfo{author}{\bibfnamefont{R.}~\bibnamefont{Fernandez-Feria}},
  \bibinfo{journal}{Phys.~Fluids} \textbf{\bibinfo{volume}{30(7)}},
  \bibinfo{pages}{2063} (\bibinfo{year}{1987}).

\bibitem[{\citenamefont{Kerkhof}(2007)}]{kerkhof_2007}
\bibinfo{author}{\bibfnamefont{P.}~\bibnamefont{Kerkhof}},
  \bibinfo{journal}{submitted to A.I.Ch.E.~Journal}  (\bibinfo{year}{2007}).

\bibitem[{\citenamefont{Struminskii}(1974)}]{struminskii_1974}
\bibinfo{author}{\bibfnamefont{V.}~\bibnamefont{Struminskii}},
  \bibinfo{journal}{Prikladnaya Mathematica i Mechnica (Applied Mathematics and
  Mechanics)} \textbf{\bibinfo{volume}{38}}, \bibinfo{pages}{203}
  (\bibinfo{year}{1974}).

\bibitem[{\citenamefont{Lorentz}(1905)}]{lorentz_1905}
\bibinfo{author}{\bibfnamefont{H.}~\bibnamefont{Lorentz}},
  \bibinfo{journal}{Proc.~Amst.~Acad.} \textbf{\bibinfo{volume}{7}},
  \bibinfo{pages}{684} (\bibinfo{year}{1905}).

\bibitem[{\citenamefont{Grad}(1960)}]{grad_1960}
\bibinfo{author}{\bibfnamefont{H.}~\bibnamefont{Grad}},
  \emph{\bibinfo{title}{Rarified Gas Dynamics}} (\bibinfo{publisher}{Pergamon,
  New York}, \bibinfo{year}{1960}), p. \bibinfo{pages}{100}.

\bibitem[{\citenamefont{Schmieleski and
  Ferziger}(1967)}]{chmieleski_ferziger_1967}
\bibinfo{author}{\bibfnamefont{R.}~\bibnamefont{Schmieleski}} \bibnamefont{and}
  \bibinfo{author}{\bibfnamefont{J.}~\bibnamefont{Ferziger}},
  \bibinfo{journal}{Phys.~Fluids} \textbf{\bibinfo{volume}{10}},
  \bibinfo{pages}{364} (\bibinfo{year}{1967}).

\bibitem[{\citenamefont{Goebel et~al.}(1976)\citenamefont{Goebel, Harris, and
  Johnson}}]{goebel_johnson_1976}
\bibinfo{author}{\bibfnamefont{C.}~\bibnamefont{Goebel}},
  \bibinfo{author}{\bibfnamefont{S.}~\bibnamefont{Harris}}, \bibnamefont{and}
  \bibinfo{author}{\bibfnamefont{E.}~\bibnamefont{Johnson}},
  \bibinfo{journal}{Phys.~Fluids} \textbf{\bibinfo{volume}{19(5)}},
  \bibinfo{pages}{627} (\bibinfo{year}{1976}).

\bibitem[{\citenamefont{Kolodner}(1957)}]{kolodner_1957}
\bibinfo{author}{\bibfnamefont{I.}~\bibnamefont{Kolodner}},
  \bibinfo{journal}{Report NYO-7980}  (\bibinfo{year}{1957}).

\bibitem[{\citenamefont{Kohler}(1948)}]{kohler_1948}
\bibinfo{author}{\bibfnamefont{M.}~\bibnamefont{Kohler}},
  \bibinfo{journal}{Zeitschr.~Physik} pp. \bibinfo{pages}{772--779}
  (\bibinfo{year}{1948}).

\bibitem[{\citenamefont{Kumar}(1966)}]{kumar_1966}
\bibinfo{author}{\bibfnamefont{K.}~\bibnamefont{Kumar}},
  \bibinfo{journal}{Ann.~Phys.~(NY)} \textbf{\bibinfo{volume}{37}},
  \bibinfo{pages}{113} (\bibinfo{year}{1966}).

\bibitem[{\citenamefont{Kremer et~al.}(2006)\citenamefont{Kremer, Bianchi, and
  Soares}}]{kremer_soares_2006}
\bibinfo{author}{\bibfnamefont{G.~M.} \bibnamefont{Kremer}},
  \bibinfo{author}{\bibfnamefont{M.~P.} \bibnamefont{Bianchi}},
  \bibnamefont{and} \bibinfo{author}{\bibfnamefont{A.~J.}
  \bibnamefont{Soares}}, \bibinfo{journal}{Phys.~Fluids}
  \textbf{\bibinfo{volume}{18}}, \bibinfo{pages}{037104}
  (\bibinfo{year}{2006}).

\bibitem[{\citenamefont{Sharipov and Kalempa}(2003)}]{sharipov_kolempa_2003}
\bibinfo{author}{\bibfnamefont{F.}~\bibnamefont{Sharipov}} \bibnamefont{and}
  \bibinfo{author}{\bibfnamefont{D.}~\bibnamefont{Kalempa}},
  \bibinfo{journal}{Phys.~Fluids} \textbf{\bibinfo{volume}{15(6)}},
  \bibinfo{pages}{1800} (\bibinfo{year}{2003}).

\bibitem[{\citenamefont{Struminskii}(1996)}]{struminskii_1996}
\bibinfo{author}{\bibfnamefont{V.}~\bibnamefont{Struminskii}},
  \bibinfo{journal}{Prikladnaya Mathematica i Mechnica (Applied Mathematics and
  Mechanics)} \textbf{\bibinfo{volume}{60}}, \bibinfo{pages}{978}
  (\bibinfo{year}{1996}).

\end{thebibliography}

\end{document}